\documentclass[11pt]{article}
\usepackage[utf8]{inputenc}
\usepackage[margin=1in]{geometry}

\usepackage{microtype}
\usepackage{graphicx}
\usepackage{subfigure}
\usepackage{booktabs} 
\usepackage{natbib}
\setcitestyle{open={(},close={)}}

\usepackage{colortbl}
\usepackage{amsmath}
\usepackage{mathtools}
\usepackage{amsthm}
\usepackage{algorithmic}
\usepackage{graphicx}
\usepackage{textcomp}
\usepackage{xcolor}
\usepackage{amsfonts}     
\usepackage{epsfig}
\usepackage{wrapfig}
\usepackage{balance}
\usepackage{url}
\usepackage{mathtools}
\usepackage{natbib}
\usepackage{multirow}
\usepackage{soul,comment}

\usepackage[textsize=tiny]{todonotes}

\usepackage[colorlinks=true,citecolor=blue]{hyperref}

\usepackage{amsmath,amsthm,amsfonts,amssymb,mathdots,array,mathrsfs,bm,bbm,stmaryrd,graphicx,subfigure,xcolor}

\usepackage{breakcites}

\usepackage[T1]{fontenc}
\usepackage{enumerate}
\usepackage{inputenc}

\usepackage{graphicx} 
\usepackage{subfigure}

\usepackage{booktabs,balance}
\usepackage{rotating}
\usepackage{boldline}
\usepackage{makecell}
\usepackage{multirow}
\usepackage{balance}

\usepackage{natbib}
\setcitestyle{open={(},close={)}}

\usepackage{algorithm} 
\usepackage{algorithmic}  
\usepackage[algo2e]{algorithm2e} 
\usepackage{enumitem}

\usepackage{xspace}
\usepackage{tabu}
\def\sys{{\tt NL2GDPR}\xspace}    
\newcommand{\para}[1]{{\vspace{2pt} \bf \noindent #1 \hspace{3pt}}}

\begin{document}

\title{\bf\Huge NL2GDPR: Automatically Develop GDPR Compliant Android Application Features from Natural Language}

\author{\vspace{0.2in}\\\textbf{Faysal Hossain Shezan, Yingjie Lao, Minlong Peng, Xin Wang, Mingming Sun}\\\\
\textbf{Ping Li} \\\\
Cognitive Computing Lab\\
Baidu Research\\
10900 NE 8th St. Bellevue, Washington 98004, USA\\
No. 10 Xibeiwang East Road, Beijing 100193, China\\\\
 \texttt{\{faysalhossain2007,laoyingjie,wwypml,sanchih.wang,sunmingming,pingli98\}@gmail.com}
}

\date{\vspace{0.2in}}
\maketitle

\begin{abstract}\vspace{0.3in}

\noindent The recent privacy leakage incidences and the more strict policy regulations demand a much higher standard of compliance for companies and mobile apps. However, such obligations also impose significant challenges on app developers for complying with these regulations that contain various perspectives, activities, and roles, especially for small companies and developers who are less experienced in this matter or with limited resources. To address these hurdles, we develop an automatic tool, {\tt NL2GDPR}, which can generate policies from natural language descriptions from the developer while also ensuring the app's functionalities are compliant with General Data Protection Regulation (GDPR). {\tt NL2GDPR} is developed by leveraging  an information extraction tool, \textbf{OIA (Open Information Annotation)}, developed by~\citet{sun2020predicate,wang2022oie} from Baidu Cognitive Computing Lab.

\vspace{0.1in}
\noindent At the core, {\tt NL2GDPR} is a privacy-centric information extraction model, appended with a GDPR policy finder and a policy generator. We perform a comprehensive study to grasp the challenges in extracting privacy-centric information and generating privacy policies, while exploiting optimizations for this specific task. With {\tt NL2GDPR}, we can achieve 92.9\%, 95.2\%, and 98.4\% accuracy in correctly identifying GDPR policies related to personal data storage, process, and share types, respectively. To the best of our knowledge, {\tt NL2GDPR} is the first tool that allows a developer to automatically generate GDPR compliant policies, with only the need of entering the natural language for describing the app features. Note that other non-GDPR-related features might be integrated with the generated features to build a complex app.

\end{abstract}

\newpage

\section{Introduction}
\label{sec:introduction}

As the technology evolves, security and privacy have emerged as important concerns for modern systems and applications~\citep{doan2021back,doan2021lira}. Due to various data breach incidents in the past years~\citep{data-breach-equifax, cambridge-analytica}, security experts and researchers have been increasingly focusing on improving the protection of user privacy.
To mitigate such risks, European Union (EU) introduced {\it General Data Protection Regulation} (GDPR) on May 25, 2018, to ensure a secure and safe standard of data usage practice~\citep{gdpr-policy}. It imposes obligations onto organizations anywhere in the world, as long as they deal with data from the people in the EU. 
For example, GDPR gives individuals the right to ask for their data to be deleted and organizations are obligated to do so. The GDPR will proceed with harsh fines against those who violate its privacy and security standards~\citep{gdpr-fines-1, gdpr-fines-2}. There are several other privacy policies (e.g., CCPA~\citep{ccpa-policy}, CDPA~\citep{cdpa-policy}, CPA~\citep{cpa-policy}) similar to GDPR that have been developed recently. In this study, we primarily focus on GDPR as it covers more population than the others.

Among all the challenges that the companies are facing to comply with GDPR policies, three main reasons stand out:  
(i) {\it Long and complex policies.} GDPR policies are very long and complex to interpret~\citep{guaman2021gdpr}. There is no simpler or less strict version for the small companies. The same GDPR policies are applied across all the companies as long as one processes data from EU citizens, which could create an excessive burden for companies with small sizes.
(ii) {\it Gap in understanding.} There are often gaps in GDPR implementations~\citep{kununka2017comparative}, where organizations tend to focus on the legal aspects, contracts, security, and data protection officers, while overlooking other key elements. 
(iii) {\it Lack of skill and awareness.} Even worse, as developers are typically not trained against GDPR policies, researchers highlighted the major causes of poorly implemented security and privacy mechanism in apps (we refer mobile application as `app') were from the problem of inexperienced, distracted, or overwhelmed developers~\citep{acar2016you}. 

Many popular mobile apps were found to violate GDPR policies specifically due to this, by improperly obtaining user permission on data collection or lacking the proper interaction with the user before collecting the user's private data. For example, previously there was no option to actively opt-out from Amazon's data collection procedures~\citep{gdpr-not-compliance}. In contrast, the Waze app displays the breakdown of the purpose of using personal data, and the user must click the `Agree' button to grant such permission to Waze. In this way, Waze complies with GDPR in terms of getting consent.

To this end, statistics show that around 91\% companies need to recruit a dedicated team or third party firm for GDPR compliance~\citep{cost-gdpr-compliance}. It is costly, and many small companies cannot even afford it. Recent works tried to reduce the gap between API documentation and the corresponding implementations~\citep{nguyen2017stitch}, which, however, did not offer much for complying with GDPR. For instance, they would not detect the requirement of `data deletion' of the apps and provide the corresponding functionalities to comply with the data deletion policy. On the other hand, existing works on policy generation (AutoPPG~\citep{yu2015autoppg} and PrivacyFlash Pro~\citep{zimmeck2021privacyflash}) mainly focused on automatically generating GDPR compliant privacy policies, without checking whether the functionalities of the apps indeed comply with GDPR.
Web-based privacy policy tools determine whether a particular website is GDPR compliant by checking the cookies~\citep{gdpr-cookie-checker-1, gdpr-cookie-checker-2}, asking template based questions \citep{gdpr-questionnaire-checker-1, gdpr-questionnaire-checker-2}. But they are only limited to suggesting what the developer should do.  Other paid tools generate the corresponding policies for the app~\citep{gdpr-paid-checker-1, gdpr-paid-checker-2}.

\newpage

In this work, we aim to automatically build GDPR compliant android features (definition of feature in Section~\ref{subsec:definition-feature-pii}) from natural language descriptions to help developers alleviate the burden of excessive efforts to comply with GDPR policies. To the best of our knowledge, this is the first work on generating policies from natural language. We follow the direction of the existing template-based tools but incorporate more flexibility with natural language as the entry, leading to more diverse and customized privacy policies across different users. Specifically, we consider the following GDPR policies -- retention, consent, privacy policy, access, deletion, third-party data sharing, and data processing security. We also ensure the functionality of the generated feature complies with GDPR. In this way, they won't need to go through the long and complex GDPR policies. Using our tool, we predict the requirements of the corresponding GDPR policies, which help the developers in finding the right applicable GDPR policies. As our tool is automatically building the GDPR compliant feature, a lack of awareness of the GDPR policies would not affect the developers. We envision our tool can be the most beneficial to smaller companies or freelance developers that do not have sufficient resources for the extra effort in complying with GDPR policies. To this end, we only consider building the simple feature in this paper. Other non-GDPR-related functionalities can also be integrated with our generated features to build larger and complex features. Note that large apps would also involve more complex and specific functionalities, which will vary across different apps.

\vspace{0.1in}
\noindent\textbf{Challenges.} There are several key challenges that we face while building the tool. 
First, since we start from natural language descriptions, unstructured description makes it difficult to detect the feature of the description, or identify the user interface (UI) element interaction on both single and multiple screens. Without accurately detecting these, it would be impossible to build an end-to-end mobile app.  
Second, various types of usage and processing of personal information (PII) in the app require compliance with different GDPR policies. Failure to detect the type of usage also likely leads to a violation. 
Third, to comply with GDPR, we need to generate a detailed description of the PII usage and privacy policies such that users of the apps can easily understand.

\vspace{0.1in}
\noindent\textbf{Our Approaches.} 
We introduce \sys to analyze the text description of the app to generate a GDPR compliant mobile feature. Our tool has three main components: (i) information extractor, (ii) GDPR policy finder, and (iii) policy generator. First, we analyze the description to extract important information by using the \textit{information extractor} component. This component is responsible for extracting UI elements, features, and page transitions. We build our information extractor with the help of a cutting-edge information extraction tool (i.e., OIA,  Open Information Annotation) and a rule-based approach (Section~\ref{sec:system-design}) with optimization and customization for our privacy-centric task. We identify all the coreference (coref)~\citep{lee2017end,fei2019end,xu2020revealing} between page transitions using a recent coref tool. Second, we investigate the GDPR policy for each of the descriptions using the proposed \textit{GDPR policy finder}. To do that, we employ a rule-based approach incorporating with OIA, Named Entity Recognizer (NER) tagger~\citep{finkel2005incorporating}, and the same coref tool to predict the required GDPR policies. We use NER to detect the entity with which the app is sharing data. Third, based on the obtained information, we use our \textit{policy generator} to create GDPR compliant privacy policies. We adapt the existing works on paraphrase generation to introduce diversity in our generated policies. We measure the quality of the generated policies (by calculating the readability scores) and select the best one to include as a privacy policy in the app. Finally, we use MIT app inventor to convert our intermediate logic into an executable apk file.

\newpage
\noindent\textbf{Contributions.} The main contributions of this work are summarized below:
\begin{itemize}    
    \item \textbf{Information Extractor.} We develop an end-to-end system to extract necessary information from natural language descriptions for generating GDPR compliant features.
    \item \textbf{Privacy Policy Generator.} We automatically incorporate the extracted information to generate a diverse set of privacy policies. To ensure the quality of the privacy policies, we also enhance the readability of the generated policies. 
    \item \textbf{Survey.} We perform a dedicated and comprehensive survey to collect natural language descriptions of 114 mobile apps.
    \item \textbf{Dataset.} We will release the entire dataset and all the generated data, including all the experimental results, app descriptions from the participants, and generated privacy policies for facilitating future advances. 
\end{itemize}

\section{Background}
\label{sec:background}
In this section, we start with introducing the terminology used in the paper, GDPR policies, and requirements, and then followed by descriptions of tools used in this paper, including OIA, coref, and NER tagger.

\subsection{GDPR Policy}  \label{sec:background:gdpr-policy-definition}

Alongside techniques for protecting the security and privacy of user data and  systems~\citep{liu2018survey,adi2018turning,he2019sensitive,lao2022identification,lao2022deepauth,yang2021robust,zhao2022integrity}, law and legal regulations are also crucial in enforcing a safe standard for practical applications. The legal requirements of GDPR policies are described in 99 different articles~\citep{gdpr-policy}. To illustrate the details, there are 173 recitals that provide further context and clarifications to those articles. In this work, we are highlighting the GDPR policies that are related to storing, processing, and sharing of PII information (definition of PII is described in Section~\ref{subsec:definition-feature-pii}), as many app owners faced huge amounts of fines due to not complying with those policies previously. We summarize the investigated GDPR policies in Table~\ref{tab:gdpr-policies} and list the details~below.

\begin{table}[!ht]
\centering
\caption{GDPR policies. Here, controller determines the purposes and means of processing personal data of data subject.}
\vspace{0.1in}
\begin{tabu}{p{0.2cm}p{2cm}p{1cm}p{11cm}}
\toprule
    \textbf{ID} &  \textbf{Policy}  & \textbf{\#Art} & \textbf{Description} \\  \toprule
    P1 & Retention & 5 & personal data should not be stored longer than requirement \\  \hline
    P2 & Consent & 6, 7 & data subject should provide consent before processing their personal data and they should have the option to withdraw their consent at any time \\  \hline
    P3 & Privacy Policy & 13  & while processing personal data, the controller should describe the purpose and the details of that in the privacy policy clearly \\  \hline
    P4 & Access & 15 & data subject has the right to request for access to their processed personal data \\  \hline
    P5 & Deletion & 17 & data subject has the right to request  erasure of their personal data \\  \hline
    P6 &  Sharing & 28  & controller is responsible for ensuring that any third-party processor should comply with GDPR \\  \hline
    P7 &  Security of Processing & 32  & the controller should take appropriate measurement to protect the security of the personal data while processing \\

\bottomrule
\end{tabu}
\label{tab:gdpr-policies}
\end{table}

\vspace{0.1in}\noindent\textbf{Retention:} If the app collects personal user data, they should not store that data for lifetime. They should state the duration of the data storage in their privacy policy. After expiring, they should remove it from their database.

\vspace{0.1in}\noindent\textbf{Consent:} App needs to get consent before collecting user data. The user should participate in interacting with the UI element (e.g., selecting a check box) before providing the consent. If the interaction is missing from the UI, the app violates the GDPR policy in taking permission~from~the~user.

\vspace{0.1in}\noindent\textbf{Privacy policy:} App should describe the purpose of collecting, processing and storing personal data in the privacy policy. They should also demonstrate the purpose of collecting personal data clearly. In the event of data sharing, they need to disclose what and how they will share the data with the third party. And the privacy policy should be straightforward. A vague privacy policy will introduce confusion among the user, which violates GDPR policies.

\vspace{0.1in}\noindent\textbf{Access:} The app may store user data for different purposes. No matter what, they should provide an option to download the stored personal data in the corresponding app. This can be downloaded in the form of a pdf, or the pdf could be directly sent to the user's requested email account.

\vspace{0.1in}\noindent\textbf{Deletion:} Article 17 is about the data deletion policy proposed by GDPR. From analyzing that article, we summarize the description of the data deletion practice according to GDPR, i.e., ``data subject has the right to request the erasure of their personal data''. The app should provide an option to delete all the stored personal data. The app owner needs to ensure that the requested data should not exist in any of their databases/systems.

\vspace{0.1in}\noindent\textbf{Sharing:} According to article 28, developer of the app needs to ensure the compliance of third-party software. They need to send personal data by ensuring the security of the communication channel. It can be achieved either by encrypting personal data or maintaining a secure communication channel. They should clearly mention their sharing regulations in the privacy policy.

\vspace{0.1in}\noindent\textbf{Security of processing:} When the app collect data from the user end, they need to ensure the security of the personal data. They should reduce the chance of data leakage during the collecting, processing, and sharing step.

\subsection{Definitions of Feature and PII}
\label{subsec:definition-feature-pii}

We define `feature' as the main functionality of a description. We take the following app description as an example: ``I want to create a `Registration' page for my app. The first page of `Registration' will have two buttons and three edittext. The edittext is used for entering the user's name, email, and password, respectively. Click the button of `sign in' to jump to the `login' page. When you click the button of `sign up', you can register your account.'' This illustrates the process of registration in the application. Here, we consider `registration' as the feature of this description.

According to GDPR, any information that can be used to identify an individual is considered as PII~\citep{pii-definition}. We list the following information as PII in this paper -- username, firstname, lastname, name, email, mail, address, country, state, zipcode, city, county, age, location, birthdate, ipaddress~\citep{pii-list, grundy2019data}. Whenever an app wants to collect these information, it needs to comply with GDPR. In other words, storing, processing, and sharing these information requires extra attention. Our tool detects the presence of these PII and stores/processes/shares the data in a GDPR compliant manner in the generated app.

\subsection{Measuring Readability} \label{sec:readability} Readability defines how easy it is to understand one's writing. The higher readability of a sentence, the easier people will be able to understand. On the other hand, when the readability is low, it demands a lot of concentration, or for most of the time, leaves a draining experience for the readers. To measure the readability of online articles (especially in medical forums), researchers have been using several different metrics~\citep{murray2019readability}. Some of the most popular formulas to calculate the readability include Dale-Chall Score~\citep{dale-chall-score}, Gunning's FOG Index~\citep{gunning-fog-score}, and Flesch Reading Ease~\citep{flesch-reading-score}, which are utilized to assess the readability level of our generated policies in this paper. 

In Table~\ref{tab:score-map-to-readability-level}, we show the score corresponding to the readability level. For example, a sentence with a 7.5 Dale-Chall score will be easily understandable by a 7th or 8th grade student. 
We expect the generated policies to be understandable by a 7th or 8th grade student. Thus, we set the readability threshold values of Dale-Chall Score, Gunning's FOG Index, and Flesch Reading Ease accordingly to reflect this readability goal.

\begin{table}[!ht]
\centering
\caption{Readability level for different score.}\vspace{0.1in}
\begin{tabular}{llll}
\toprule
   \textbf{Readability Level} & \textbf{Dale-Chall} &  \textbf{FOG}  & \textbf{Flesch}   \\  \toprule
    4th-grade student or lower & 4.9 or lower     &  0-5   &  90-100  \\  \hline
    5th or 6th-grade student   & 5.0-5.9          &  6     &  80-89    \\  \hline
  \cellcolor{gray!30}7th or 8th-grade student   &  \cellcolor{gray!30}6.0-6.9          &   \cellcolor{gray!30}7-8   &  \cellcolor{gray!30}60-79   \\  \hline
    9th or 10th-grade student  & 7.0-7.9          &  9-12  &  50-59     \\  \hline
    11th or 12th-grade student & 8.0-8.9          &  9-12  &  50-59     \\  \hline
    College student            & 9.0-9.9          &  13-16 &  30-49    \\  \hline
    Graduate student           & 10.0 or high   &  17    &  0-29    \\  

\bottomrule
\end{tabular}
\label{tab:score-map-to-readability-level}
\end{table}

\subsection{OIA Tool} \label{subsec:background:oia-tool}

\begin{figure}[!ht]
\centerline{\includegraphics[width=4.2in]{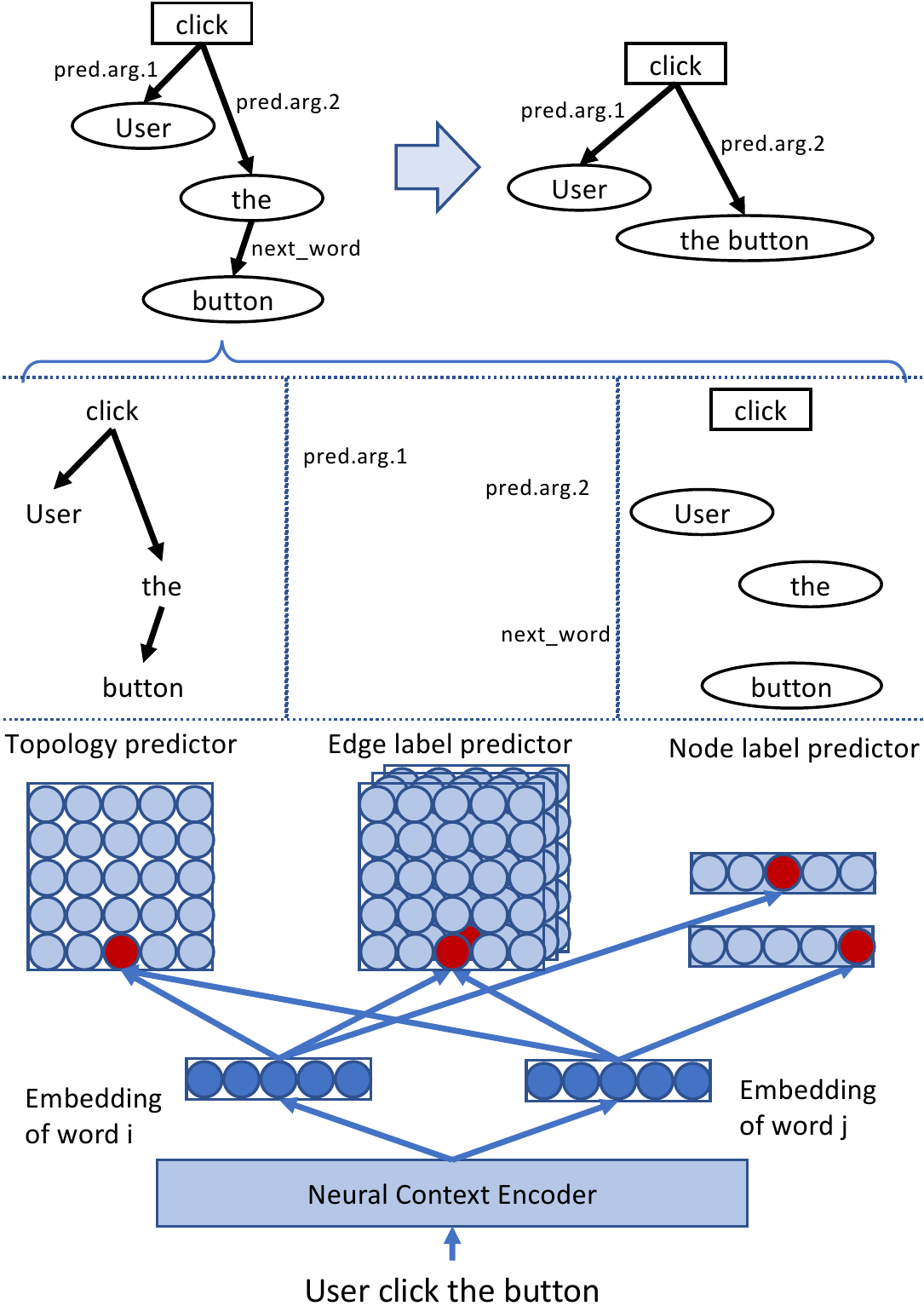}}
\caption{The process of the neural predictor converts a sentence to an OIA graph.}
\label{fig-oia}
\end{figure}

Open Information Annotation (OIA)~\citep{sun2020predicate,wang2022oie} is the recently proposed approach for building OIE systems. It represents all the information in a sentence into a predicate/function-argument and expresses them in a graph. The OIA graph is generated with a cutting-edge neural predictor. As shown in Figure~\ref{fig-oia}, the input sentence is encoded with  BERT~\citep{devlin2019bert}. As a result, each word is represented with a word embedding. Then three neural components predict the topology, edge label, and node label of the graph, respectively.

Among these, the node label predictor is a neural multi-class classifier that generates a tag for each word node based on the word embeddings. For instance, `noun' nodes (ellipse in Figure~\ref{fig-oia}) represent for entities. The `event' nodes (rectangle in Figure~\ref{fig-oia}) and their arguments represent interactions. The `prepositional' nodes connect the event and its attributes (e.g., time, location, manner, etc.).
The topology predictor is a matrix predicted based on the word embeddings. The element at row $i$ and column $j$ is used to predict whether there is a directed edge between word $i$ and word $j$. 

\newpage

The edge label predictor is a tensor predicted based on the word embeddings and used to select labels for edges between word pairs. A typical edge label like `pred.arg.x' shows the tail node is the $x$th arguments of the head node. For the event node, the first argument and is the subject, and the second argument is the object. For the prepositional nodes, the first argument is the event, and the second argument is the attribute of the event.

The results of these predictors are merged together as the directed graph with word-level nodes. Then the tool needs to merge the word-level graph (upper left in Figure~\ref{fig-oia}) into a phrase-level graph whose nodes contain descriptions of components or interactions with more than one words (upper right in Figure~\ref{fig-oia}). In the word-level graph, there are two types of edges, i.e., intra-phrase and inter-phrase edges. Intra-phrase edges (edges with label `next\_word') are the basis of graph structure changing. Word-level nodes connected by `next\_word' are merged into phrase-level nodes. In contrast, inter-phrase edges (edges with other labels) remain unchanged. Finally, the tool can get the OIA graphs that describe the interactions or system actions.

\newpage

The OIA graph can be used to efficiently and effectively harvest `event' nodes and their argument sub-trees as the information extraction result. Evaluating on the recently widely-used open information extraction benchmark Re-OIE2016~\citep{zhan2020span}, OIA achieves the best result and improves accuracy by about 2\% from the prior best result reported in~\citet{ro2020multi}.

\subsection{Coref Tool}\label{subsec:background:coref-tool}

In this paper, we use an end-to-end neural network-based coreference (coref) resolution method~\citep{lee2017end,fei2019end,xu2020revealing} to resolve mentions (i.e., words or phrases that refer to entities) and their antecedents. Coref aims to identify all mentions referring to the same entity. The input could be a text document, while the output is several clusters of mentions. Mentions in the same cluster refer to the same real-world entity. We use the coref tool provided by~\citet{xu2020revealing,joshi2020spanbert} to implement this step along with adaptations for the privacy-centric task. 

\begin{figure*}[!ht]
\centerline{\includegraphics[width=5in]{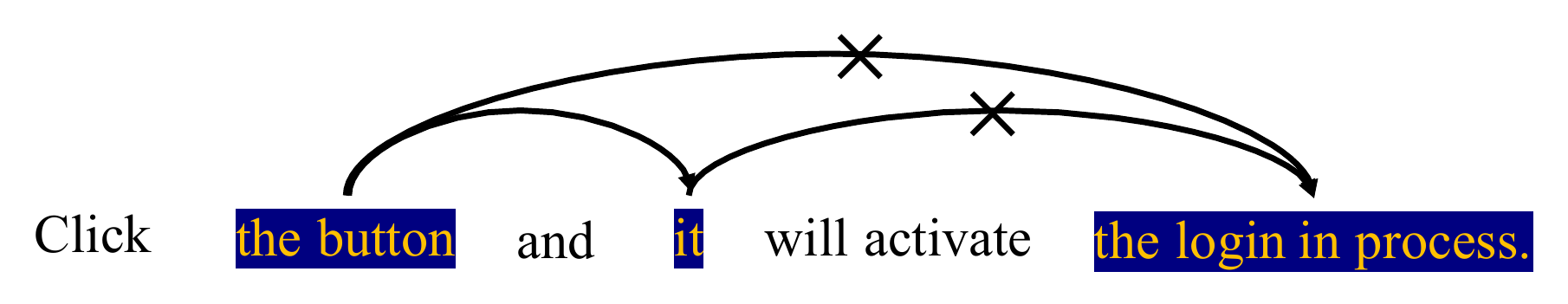}}
\caption{Mention detection and coreference resolution.}
\label{fig-cr}
\end{figure*}

Specifically, the word representations are generated with BERT~\citep{devlin2019bert} in the tool~\citep{xu2020revealing,joshi2020spanbert}. All continuous words can be a candidate span. Span representations are generated by composing representations of containing words. Then, the tool predicts whether a span is an entity mention based on its representation. After extracting the entity mentions, the tool predicts whether a mention pair represents the same entity. It scores each pair based on the span representations and considers pairs with scores higher than a threshold as coreference.
Taking the sentence depicted in Figure \ref{fig-cr} as an example, ``the button'', ``it'' and ``the login process'' will be recognized as mentions by the coreference resolution tool. In addition, ``the button'' and ``it'' will be recognized as two co-referred mentions corresponding to the same entity. 

\subsection{NER Tagger} Named Entity Recognizer (NER) tagger indicates the name of three classes, i.e., person, organization, and location, in a sequence of words. In this work, we used the Stanford NER tagger~\citep{finkel2005incorporating} to identify the organization name with whom the app will share PII data. We mark all the organizations as third-party. For example, in the description of ``we will share your data with Facebook'', the NER tagger will identify `Facebook' as an organization.

\section{Related Works}
\label{sec:related-work}

This section outlines the related works, including automatic code generator and GDPR privacy policy analysis, and highlights the differences and our novelty.

\subsection{Automatic Code Generation}

In the literature, researchers used several approaches for automatic code generation, such as leveraging graphical user interfaces to generate code~\citep{beltramelli2018pix2code}, automatically converting code from description to shell script~\citep{lin2017program}, generating free form java queries~\citep{gvero2015synthesizing}, introducing plugins to help developers to implement APIs correctly~\citep{nguyen2017stitch}, and generating attacks specific to online app generators~\citep{oltrogge2018rise}. At the same time, attempts have also been made to use natural descriptions to create single-page android applications with limited functionalities~\citep{hasan2021text2app}. In contrast, we are neither limiting ourselves to a single page nor simple functionalities in our work. Our function involves operations with personal data processing and storing. It covers from how the data will be transmitted over the network to how long it will be stored in the database. Besides, we are converting long descriptions into code that involves multiple pages and several interactions in the UI elements.

\subsection{GDPR Privacy Policy Analysis} Identifying GDPR violence in the existing apps is a hot research topic. Researchers are working on determining applications violating GDPR policies by performing measurement analysis~\citep{ferrara2018static, guaman2021gdpr}, characterizing the impact of GDPR~\citep{linden2020privacy} in online privacy policies, which have already led to the findings of many inconsistencies in policies~\citep{zimmeck2019maps, zimmeck2014privee}. Some applications' privacy policies were very vague in that one part of the policy indicated not sharing any data with third parties, whereas in other parts, the application developer described the process of sharing some portions of the collected data with third parties~\citep{andow2019policylint}. Prior works have also investigated the source code of popular android apps and IoT platforms to find the information leakages via over-sharing of information through third parties~\citep{nan2018finding, shezan2020tkperm, shezan2020verhealth}. Among these prior works, many target medical applications as those dealing with a lot of sensitive personal health information~\citep{fan2020empirical, kammuller2019designing}. In sum, the existing works are focusing more on improving the transparency of personal information usage~\citep{liu2018towards, miao2014privacyinformer, tesfay2018read, utz2019informed, waddell2016make}, which typically aim at generating privacy policies from source codes~\citep{yu2015autoppg, zimmeck2021privacyflash}. In contrast, our tool generates GDPR compliant privacy policies from the natural language descriptions, while ensuring the app's functionality is compliant with GDPR and disclosing all the private data processing and collection.

\subsection{Information Extraction}

Information extraction is a task to extract desired information in the form $(m_A, relation, m_B)$ from sentences, where $m_A$ and $m_B$ are entity mentions in a sentence, and $(m_A, relation, m_B)$ indicates there is a $relation$ between mention $A$ and mention $B$. Traditional information extraction methods~\citep{mintz2009distant,zhang2021readsre,wang2022deep,zhang2022end} need a pre-defined relation set and the corresponding set of positive examples $\{(m_A, relation, m_B)\}$ for each relation to train a relation extractor, which is costly and time-consuming. 

An alternative approach is Open Information Extraction (OIE), where $relation$ is expressed by natural language phrases, so it does not need a pre-defined relation set ~\citep{sun2018logician,sun2018logician_orator,cui2018neural,stanovsky2018supervised,liu2020advantage,liu2020extracting,ro2020multi,zhan2020span}. 
However, traditional OIE system develops a specific machine for every single task. Without a unified standard, the OIE systems are isolated from each other. It is hard to find an OIE system exactly matching the requirement of our privacy-centric task.

Recently, a concept called Open Information eXpression (OIX) was proposed by~\citet{sun2020predicate,wang2022oie} to address the adaptability issue of OIE systems. The idea of OIX is to introduce an intermediate layer between the language and OIE. \citet{sun2020predicate} proposed a standard, called Open Information Annotation (OIA), to implement OIX, and introduced a rule system to implement OIA. In the following, \citet{wang2022oie} introduced a more efficient neural-network-based model to implement OIA, and achieve new SOTAs on four OIE tasks with simple rules.

As discussed in Section~\ref{subsec:background:oia-tool}, the OIA graph contains all information so that users can extract the facts of their interest by simple rules, eliminating the necessity of building different OIE systems for different tasks. In this paper, we extract the tailored information of UI elements, features, and page transitions from the OIA with optimizations for our task.

\section{Data Collection}
\label{sec:data-collection}

We collect natural language descriptions of 114 apps by performing a survey. The detailed information about the survey are listed in Section~\ref{appendix:survey-questionnaire}. Each of the participants is required to have at least two years of development and extensive user experience on mobile applications. Such expertise ensures that the participants have the necessary knowledge of the mobile applications and processing of different user data for this survey. We perform the survey in two phases. In the first phase, we ask five participants to participate. We collect 66 app descriptions which contain 258 sentences  (containing 23 features) from them. In the second phase, we collect the data of 48 apps (containing 396 sentences) from different participants, while ensuring each app is described by at least three participants. In the study, we ask the participants to describe each app in natural language. We provide them a guideline with a list of information that they need to use in their description. Our information contains (i) feature list, (ii) UI elements, and (iii) PII data.  In total, we have 24 features, 17 UI elements, and 17 PII data. 
To help them write high-quality descriptions, we have included a few examples of different apps along with the corresponding screenshots. It will also help them to understand the guideline for writing the description correctly. It took only 15-17 mins on average to describe each app.

\subsection{Survey Questionnaire}
\label{appendix:survey-questionnaire}

In the survey, we show the lists of features, UI elements, PII data that the participants can use.  For each feature, we provide one to three screenshots for the description writers, in order to help them understand the feature. Then, the writers can describe the details of the feature, e.g., what UI element it should have, what kind of data the user should provide, what type of action will be activated, etc. In the following, we list the survey questionnaire that we present to the participants.

\vspace{0.1in}\noindent\textbf{Features/Actions:} login, logout, register, forget password, status updates, share, user profile, third-party integrations, in-app advertisement, connections, newsfeed, social authorization, post creation, post likes, notifications, analytics. 
To learn more about actions, visit here: \url{{https://www.businessofapps.com/insights/social-\\networking-app-features-that-make-it-happen}}.

\vspace{0.1in}\noindent\textbf{UI Elements:} TextView, EditText, Button, ImageButton, ToggleButton, RadioButton, RadioGroup, CheckBox, AutoCompleteTextView, ProgressBar, Spinner, TimePicker, DatePicker, SeekBar, AlertDialog, Switch, RatingBar

To learn more about UI elements, please visit: \url{data-flair.training/blogs/android-ui-controls/}.
 
\vspace{0.1in}\noindent\textbf{Events:} press, long pressed

To learn more about events, please visit: \url{developer.android.com/guide/topics/ui/ui-events.}
 
\vspace{0.1in}\noindent\textbf{Resource:} image, text, video

\vspace{0.1in}\noindent\textbf{Data:} username, firstname, lastname, name, email, mail, address, country, state, zipcode, city, county, age, location, birthdate, ipaddress

To learn more about functionalities of various mobile applications, please visit the following links:
\begin{itemize}[topsep=1pt, itemsep=1pt, partopsep=1pt, parsep=1pt]
    \item Facebook: \url{https://play.google.com/store/apps/details?id=com.facebook.katana} 
    \item Instagram: \url{https://play.google.com/store/apps/details?id=com.instagram.android}
    \item LinkedIn: \url{https://play.google.com/store/apps/details?id=com.linkedin.android}
    \item Snapchat: \url{https://play.google.com/store/apps/details?id=com.snapchat.android}
\end{itemize}

You (i.e., participants) need to select features that you want to implement and describe the functionalities of that feature using the attributes listed above.

\begin{figure}[h!]
    \centering
    \subfigure[(a) Snapchat.]{\includegraphics[height=8cm]{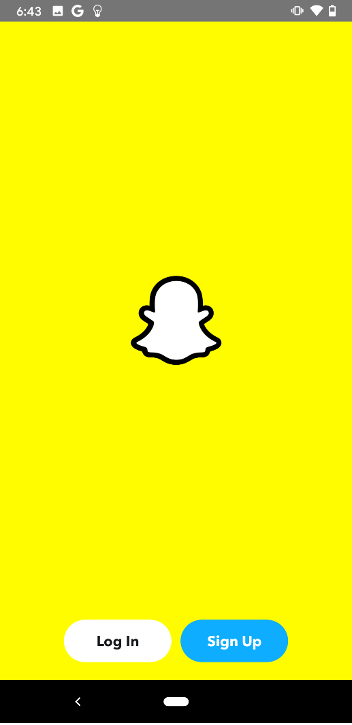}}\label{fig:log_in}
    \hspace{3cm}
    \subfigure[(b) Peacocktv app.]{\includegraphics[height=8cm]{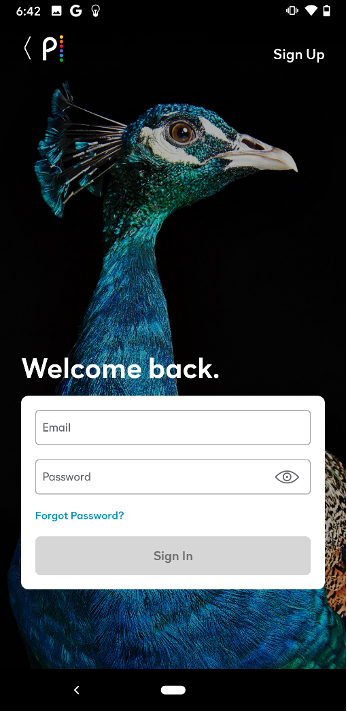}}\label{fig:sign_in}
    \caption{App screenshot for login.}
    \label{fig:full:login}
\end{figure}

\vspace{0.1in}\noindent\textbf{Example \#1} Figure~\ref{fig:full:login}(a) presents the screenshot of the `introduction page' of a mobile application downloaded from Google Playstore. (\url{https://play.google.com/store/apps/details?id=com.snapchat.android}). Here, we can see one button for `Log In', another button for `Sign Up', and also the logo of Snapchat in the imageview. Thus, functionality can be described as:
\textit{I want to develop an `introduction' page for my mobile application. On that page, there will be two buttons; one button is for `Log In' and the other button is for `Sign Up'. When the user clicks the `Log In' button, it will take to the `Log In' page. Whereas, if the user clicks the `Sign Up' button, it will take her to the `User Registration' page. There will be an imageview that will show the application logo.}

\vspace{0.1in}\noindent\textbf{Example \#2}
Figure~\ref{fig:full:login}(b) presents the screenshot of the 'Sign In' page of a mobile application downloaded from Google Playstore (\url{https://play.google.com/store/apps/details?id=com.peacocktv.peacockandroid}). On this page, we can see the functionalities of how the user can log in to use this application. Thus, functionality can be described as below:
\textit{I want to build the `Sign In' page of my application. On this page, there will be two edittexts. One edittext is for `email,' and the other edittext is for `password'. Once the user clicks the `Sign In' button, it will send the user's email and password to the server. There will be another button `Forget Password, ' which will take the user to the `forget password' page. On the top, there will be a button as `Sign Up'. When the user clicks this button, it will take to the 'registration' page.}

\section{System Design and Implementation}
\label{sec:system-design}

\begin{figure*}[!ht]
\centerline{\includegraphics[width=\textwidth]{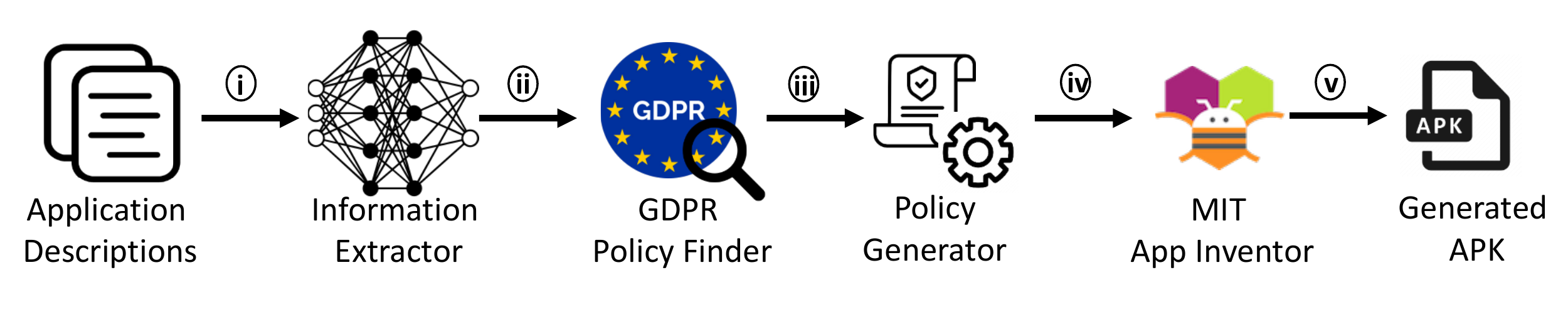}}

\vspace{-0.15in}

\caption{System architecture of \sys.}
\label{fig-system-architecture}
\end{figure*}

The system architecture of \sys is shown in Figure~\ref{fig-system-architecture}. Our system consists of three key components: (i) Information Extractor, (ii) GDPR Policy Finder, and (iii) Policy Generator. As described in Section~\ref{sec:data-collection}, We select 66 apps from phase 1 to build the component's logic of our tool. Specifically, we devise the rules for each component's detection mechanisms by analyzing these 66 app descriptions as the training data. In the following, we discuss the detailed implementations of these components.

\subsection{Information Extractor}  \label{sec:system_design:information-extractor} 

\begin{figure}[b!]

\vspace{-0.1in}

\centering
\includegraphics[width=2.8in]{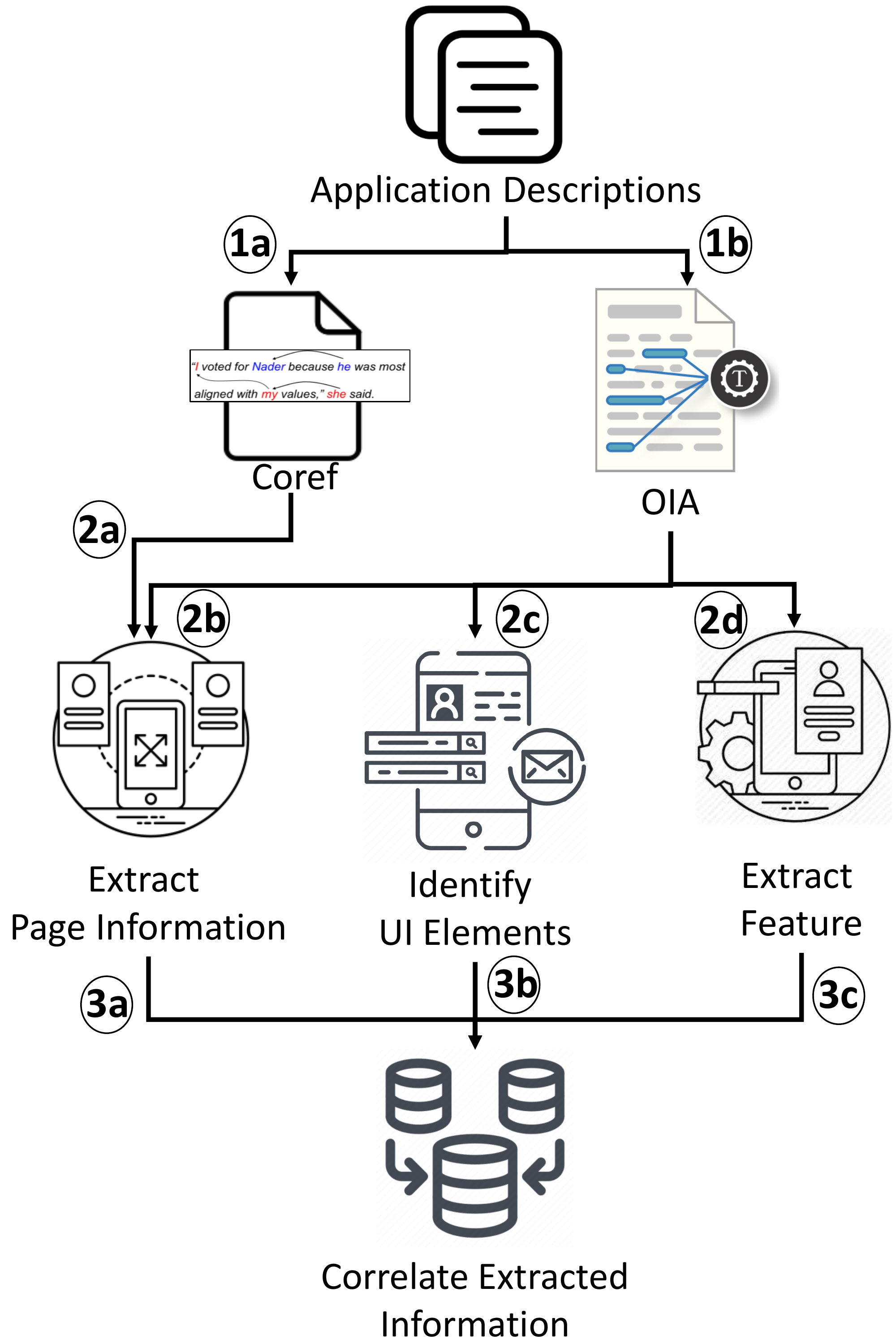}
\caption{Extracting page, UI, and feature information using the proposed Information Extractor.}
\label{fig-information-extractor-architecture}\vspace{-0.1in}
\end{figure}

To build an app, we need three types of information: page transition, UI element, and feature. We collect these information using three different sub-components of the information extractor component: Page Information Extractor, UI Information Extractor, and Feature Information Extractor. The complete process is described in Figure~\ref{fig-information-extractor-architecture}.

Existing parsers (e.g., dependency parser) require craft rules based on their results. That means we have to read the descriptions and summarize expression habits to bridge the parsing results and the information of interest. Specifically, (i) to recognize feature names, we need to merge all kinds of nominal syntactic tags with their decorative sub-graph by rules. Alternatively, we can train a recognition model with enough labeled data. (ii) To recognize the description of user interactions, we need to merge verbal phrases and analyze the semantic roles of verbal words to avoid decorative verbs (e.g., `login' in `login button'). (iii) Finally, we must coordinate the results of (i) and (ii) to eliminate conflicts and ensure a corresponding relationship between user interactions and app components. These processes will involve data annotation, rule crafting, and analysis of part-of-description, syntax, and semantics. In our task, we have to carefully design the implementations and adapt tools with customization for these steps, as any step might bring errors and hurt the performance of the entire process. In the following, we discuss the details of the three sub-components.

\subsubsection{Page Information Extractor}  \label{sec:system_design:page-information-extractor}

In each sentence of the description, participants are required to describe different pages and the related transitions. For each page, all the UI elements and functionalities of these UI elements are described. For example, ``clicking `my profile' button in `home' page will take the user to `profile' page. In `profile' page there will be firstname textview, lastname textview.'' We can observe how the transition happened between `home' and `profile' pages. Here, we mark the `profile' as the `current' page and `home' as the `transition' page. To capture all the UI elements on a single page, we need to identify the page name (both current and transition). We illustrate our algorithm for finding the page information in Algorithm~\ref{algo:page-info-desc}. As we discussed in Section~\ref{subsec:background:oia-tool}, OIA graph parser is designed for open information extraction. We can directly harvest user interaction from the `event' node and component description from the `noun' node. First, we check whether the `event' node represented transition or not. To decide this, we investigate the presence of transition keywords in the sentence. Based on our observation of the training set, we create the transition keyword list. This list contains the following keywords, $transitionKeywords$  = \{`jump', `return', `take', `go', `transition', `trigger'\}.  

One problem in OIA is that the mentions extracted facts may be mentions of pronouns like ``it'', ``that'', ``they'', which do not involve the information about the referred real entity like ``the registration page'', ``the home page''. To address this issue, we adapt the coref tool described in~\citet{xu2020revealing} to resolve the pronouns to the real entity.

\begin{algorithm}[!ht]
    \SetAlgoLined
    
    \textbf{Input:} description \\
    \textbf{Output:} pageInfo \\
    
    pageInfo = list()\;
    
     \ForEach{$sentence$ $\in$ $description$}{
         evNodes = `event' nodes of sentence\;
         
         nNodes = `noun' nodes of sentence\;
         
         transFlg = False\;
         
         \If{evNode contains transitionKeywords}{
            transFlg = True\;
         }
         \For{$nNode$ $\in$ $nNodes$}{
            corefInfo = findCorefInfo(nNode)\;
            
            \If{corefInfo $!=$ None}{
                pageName = corefInfo\;
            }
            \Else{
                pageName = extractByNode(nNode)\;
            }
            infoList = addPageName(infoList, pageName)\;
         }
         \If{transFlg == True $\&$ len(infoList) >= 2} {
            currentPageName = infoList[0]\;
            
            transitionPageName = infoList[1:]\;
         }
         \ElseIf{transFlg == True}{
            transitionPageName = infoList[1:]\;
         }
         \Else{
            \If{len(infoList) > 0}{
                currentPageName = infoList[0]\;
            }
         }
        pageInfo.append(set(currentPageName, transitionPageName))\;
     }
    \Return pageInfo\;
    
     \caption{Extracting page information}
     \label{algo:page-info-desc}
\end{algorithm}

For every noun node, we check the coref information. In the `findCorefInfo()' of Algorithm~\ref{algo:page-info-desc}, we traverse the full sentence to detect whether the corresponding noun node has any coref page name. If it is present, we assign that information as $pageName$. Otherwise, we extract the page name by using our rule-based approach in $extractByNode$ of Algorithm~\ref{algo:page-info-desc}. In this approach, we search for the node which contains `page' word to determine page name. For the sentences with transition functionality available, the number of extracted pages became more than two. In those cases, we extract the first page name information as the `current' page and all the later information as the `transition' page information. Since we can expect the current page name to be mentioned before the transition page by a developer, which is also consistent with our observations of our collected descriptions, it performs well by selecting the first page information as current page information.

\newpage

\subsubsection{UI Information Extractor} \label{sec:system_design:ui-information-extractor} 
\vspace{-0.05in}

Each description includes multiple sentences for describing the detailed interaction of the UI elements. From each sentence, we search the presence of UI elements by using a keyword-based approach. While performing the survey, we presented the participants with a list of UI elements. By analyzing the training data (Section~\ref{sec:system-design}), we observe that we can use the keyword list to collect all the UI elements in a sentence.

\newpage

In the survey, we show a list of UI elements that the participant is allowed to use while describing the app. Our full list of UI elements are: 

   [`TextView', `EditText', `Button',  `ImageButton', 
  `ToggleButton', `RadioButton', `RadioGroup',     
  `CheckBox', `AutoCompleteTextView', `ProgressBar', 
  `Spinner', `TimePicker', `DatePicker', 
  `SeekBar',  `AlertDialog', `Switch', 
  `switchbutton', `RatingBar', `Map', `RadioButtonControl']. From our investigation, we find that the participant use the UI elements from this list. So, we select this list to find the presence of UI elements by investigating the exact match.  
  
Once we find an exact match with the listed keyword, we detect that as a UI element. Using the page detector component, we determine the name of the page where the UI elements are located.

\subsubsection{Feature Information Extractor}  \label{sec:system_design:feature-information-extractor}

We predict the feature information in three steps. First, using the OIA graph, we detect the `noun' node in the sentences. Second, we use our rule-based approach to find the presence of the listed keywords in the corresponding sentences.
In our feature extractor component, we use the keywords listed in Table~\ref{tab:feature-extractor} to find the corresponding feature in the sentence. From our investigation, we build this keyword-function mappings. For example, whenever we find the representative keywords for `registration' (or `sign up') in the sentence, we mark that sentence as representing `registration'~feature.

\begin{table}[h!]


\centering
\small
\caption{Rule-based approach for finding feature information.}
\begin{tabular}{ll}
\toprule
     \textbf{Name}  & \textbf{Keywords} \\  \toprule
    registration & `registration', `sign up' \\ \hline
    user profile & `user profile' \\ \hline
    status updates & `status updates' \\ \hline
    news feed & `news feed' \\ \hline
    home feed & `home feed' \\ \hline
    comments & `comments' \\ \hline
    address book & `address book' \\ \hline
    login & `login', `sign in' \\ \hline
    change password & `change password', 'forgot password' \\ \hline
    people nearby & `people nearby' \\ \hline
    third-party integrations & `third-party integrations', `third-party', `Third-Party Data Sharing' \\ \hline
    chat with friends & `chat with friends',  `user friends list', `process request of creating friendship'  \\ \hline
    add new friends & `add new friends' \\ \hline
    emoticon input & `emoticon input' \\ \hline
    user status & `user status' \\ \hline
    blog writing & `blog writing' \\ \hline
    product scan & `product scan' \\ \hline
    news recommendation & `news recommendation' \\ \hline
    search for people nearby & `search for people nearby', `people nearby' \\ \hline
    app purchase & `app purchase' \\ \hline
    share & `share' \\ \hline
    review & `review' \\ \hline
    notes & `note' \\ \hline
    create new post & `create new post', `new post', `write post', `create a new post', `posting' \\ 
\bottomrule
\end{tabular}
\label{tab:feature-extractor}
 \vspace{0.2in}
\end{table}

When we find an exact match of the keywords, we mark the sentence exhibiting the corresponding feature (see the mapping in Table~\ref{tab:feature-extractor}). Third, we count the frequency of the matched feature inside a description.  Since not all the sentences indicate the feature information and the feature information may spread across multiple sentences, we aggregate all the information extracted to infer the feature of an app description. Our intuition is that the name of the final feature will appear the most frequently inside a description. Therefore, we consider the feature that is predicted the most in the sentences as the final feature of the description. For example, if the keyword of `registration' shows up in the sentences of a description twice while the keyword of `login' appears only once, we mark that description as `registration'.

\subsection{GDPR Policy Finder} \label{sec:system_design:gdpr-policy-finder} After extracting all the features and page-related information (using the proposed information extractor), we investigate the required GDPR policies for each of the descriptions using Algorithm~\ref{algo:gdpr-policy-desc}.

\begin{algorithm}[!ht]
    \SetAlgoLined
    \textbf{Input:} description, featureInfo \\
    \textbf{Output:} typGDPR \\
    
    typGDPRList = list() \;
    
   \ForEach{$sentence$ $\in$ $description$}{
        eNodes=getRepresentativeEventNodes($sentence$)\;
        
        nNodes = `noun' nodes of $sentence$\;
        
        \If{doesContainPII(nNodes)}{
            evType = detectType(eNodes) \;
            
            tg = map(evType, featureInfo) \;
            
            typGDPRList.append(tg) \;
        }
    }
    typGDPR = unique(typGDPRList) \;
    
    \Return  typGDPR \;
     \caption{Identifying GDPR policy}
     \label{algo:gdpr-policy-desc}
\end{algorithm}

\begin{figure}[h!]


\centerline{\includegraphics[width=4in]{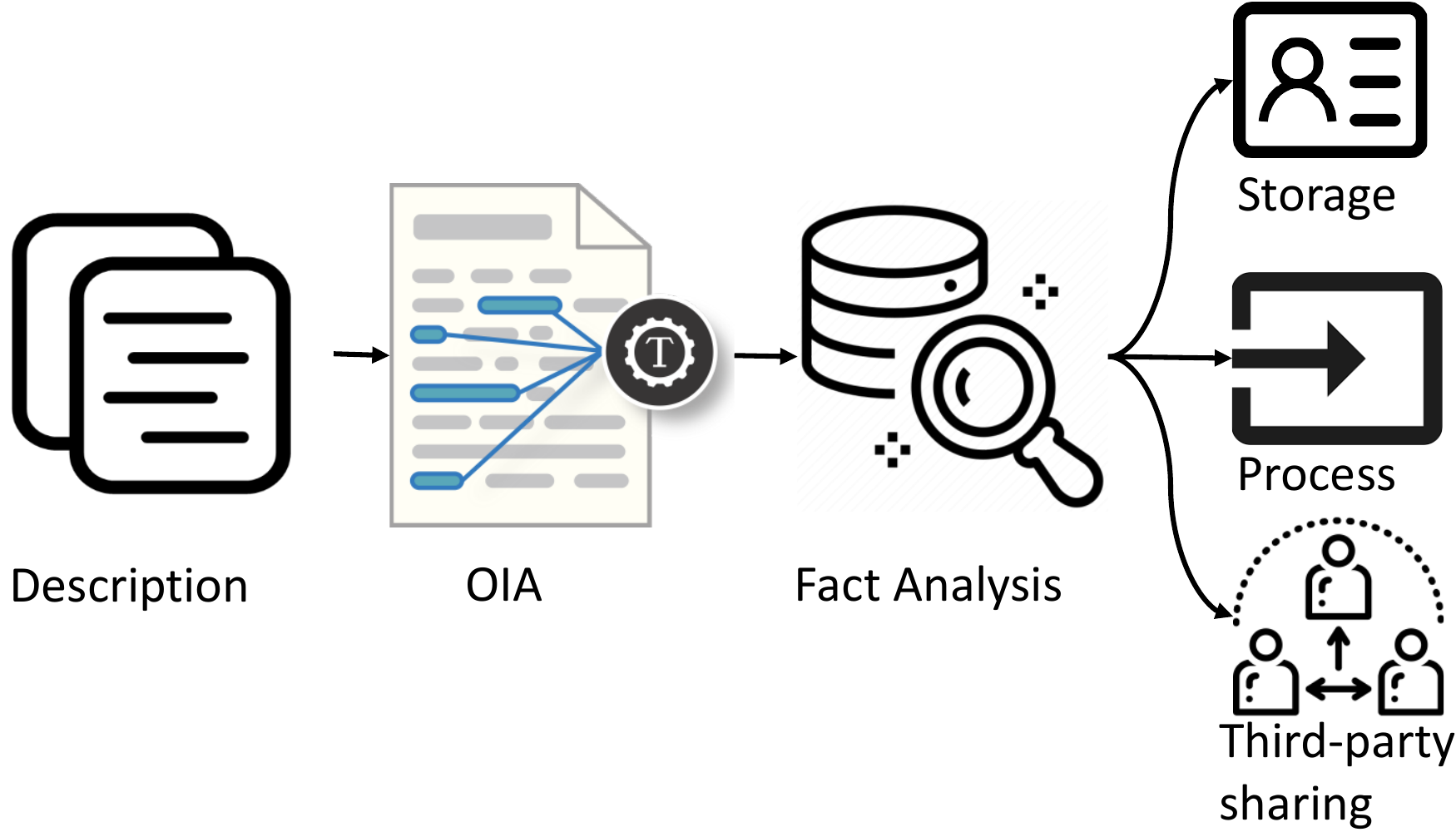}}


\caption{Using OIA to detect the required GDPR policy from description. }
\label{fig-gdpr-policy-finder}
\end{figure}

We divide this process into three parts: (i) identify PII information, (ii) identify usage of data, and (iii) generate intermediate representation of the required GDPR policy. The complete process is shown in Figure~\ref{fig-gdpr-policy-finder}. We describe each of them in the following.

\subsubsection{Identify PII Information} GDPR policies are applicable when the app uses PII. Therefore, we need to identify whether the given description contains any PII or not. First, we extract all the noun nodes using OIA. From our analysis, we observe that PII locates in the noun nodes all the time. Then, we use our keyword-based approach to find the exact match of the PII keywords. We also show a list of allowable PII that the participant can use to describe the app description. We build a PII mapping keyword set (Table~\ref{tab:pii-extractor}) to find the PII information in the description. In Table~\ref{tab:pii-extractor}, we can observe the PII list and the corresponding keywords for finding each PII. For example, we search `username' and `uname' to find all the usernames inside a sentence. In Algorithm~\ref{algo:gdpr-policy-desc}, `doesContainPII()' returns \textit{True}, whenever there is PII information available. We collect all the PII information from a description in this way.

\begin{table}[t]
\centering
\caption{Rule-based approach for extracting PII information.}
\begin{tabular}{ll}
\toprule
\textbf{PII}  & \textbf{Keywords} \\  \toprule
username & `username', `uname' \\ \hline
firstname & `firstname', `fname' \\ \hline
lastname & `lastname', `lname' \\ \hline
name     & `name', `fullname' \\ \hline
email    & `email', `mail' \\ \hline
address  & `address' \\ \hline
country  & `country' \\ \hline
state    & `state' \\ \hline
zipcode  & `zipcode', `zip' \\ \hline
city     & `city' \\ \hline
county   & `county' \\ \hline
age      & `age' \\ \hline
location & `location', `loc' \\ \hline
birthdate & `birthdate', `bday', `dob' \\ \hline
ipaddress & `ipaddress', `ip', `mac' \\ \hline
password  & `password', `pass', `pwd' \\ \hline
userinformation & `userinformation' \\ 
\bottomrule
\end{tabular}
\label{tab:pii-extractor}
\end{table}

\subsubsection{Identify Usage of Data} Once we detect PII, then we start to investigate how that information is being used. From our investigation on the training sample, we find that there could be three different types of usage -- storage, process, and third-party sharing. Sometimes, the app will store the user information (e.g., user credentials during registration). We define that as `data storage'. For completing user requests, the app may need to use user information (e.g., matching username and password during login). We define such usage as `data process'. For improving user experience  (e.g., sharing a user email address with the product manufacturer for advertisement purposes), the app may share user data with third parties. We denote such sharing as `third-party data sharing'. Let's consider the following two examples for better illustration.

\vspace{0.3in}

\fbox{\centering
\begin{minipage}{40em}
\label{minipage:example:gdpr}

\underline{Example \#1:} `login' page, user can provide her username in edittext and password in another edittext. Clicking the `login' button will take the user to the `home' page. `sign up' takes the user to the registration page.\\
\underline{Example \#2:} `register' page, user provides username in the first edittext. The second edittext user provides `first name', third edittext user provides `last name', fourth edittext user provides `password', fifth edittext user confirms `password' again. Sixth edittext user provides age. A button at the bottom which is `register' will store all the information to the server.
\end{minipage}
}

\vspace{0.2in}

In the first example, the app does not store any personal information. It only sends user credentials to the server for further user input validations. In this case, the app needs to ensure a secure way of communication with the server. Because according to GDPR, all server-client communication of personal data should happen in a secure channel and follow proper encryption protocol (according to P7). Whereas in the second example, the app stores personal information (such as first name, last name, password, etc.). Here, the app needs to comply with GDPR policies (according to P1-P5, P7).

We illustrate the rule-based approach in Table~\ref{tab:gdpr-policy-finder} where we use it for mapping event types and feature information with the type of data usage in Algorithm~\ref{algo:gdpr-policy-desc}.  For example, our information extractor predicts the second example (in code listing~\ref{minipage:example:gdpr}) as `registration' feature. We incorporate that information and look for which event keywords (listed in Table~\ref{tab:gdpr-policy-finder}) are presented in the corresponding sentence. From analyzing the sentences, we find the `store' keyword, which indicates `storage' type data usage. By following the same approach, we compute the type of data usage in each sentence. Ultimately, we compute all the unique data usage types to determine the final set of data usage types, which determines the required set of GDPR policies. 

\begin{table}[h]
\centering
\small
\caption{Rule-based approach for finding GDPR policy.}
\begin{tabu}{p{2cm}p{2.8cm}p{7cm}p{3cm}}
\toprule
     \textbf{Type}  & \textbf{Events} & \textbf{Features}  & \textbf{Rules} \\  \toprule
     Storage & Store, Save, Upload, Input, Register, Create, Record  & `registration', `user profile', `status updates’, `comments’, `address book’, `change password’, `user status’, `blog writing’, `add new friends', `notes', `create new post' & <feature>: { <event> <PII>} \\  \hline
     Process & Show, View, Display, Exhibit  & `news feed’, `home feed', `summary of the day’, `comments’, `address book’, `login’, `people nearby’, `emoticon input', `product scan', `chat with friends’, `user friends list’, `search for people nearby’, `app purchase’, `news recommendation', `share’, `review'
 & <feature>: { <event> <PII>} \\  \hline
      Third-party Sharing & Share, Send  & `third-party integrations’, `app purchase’, `share’, advertisement & <feature>: { <event> <PII> <party (identified by NER~\citep{finkel2005incorporating})>} \\ 
\bottomrule
\end{tabu}
\label{tab:gdpr-policy-finder}
\end{table}

\subsubsection{Generating Intermediate Representation} \label{subsec:generating-intermediate-representation}
Once we find all the information from the other two sub-components, we map these to the corresponding GDPR policies based on Table~\ref{tab:map-gdpr-policies}. The table shows how each of the GDPR policies is linked to the corresponding data usage type. For example, according to P2, the user should grant permission before the app can use the PII. To make the app compliant with P2, we introduce `consent' functionality whenever the app needs to take input from the user-end. This policy is also applicable where the app shares PII with other third parties. So in our generated app, we implement the consent also whenever the app is sharing data with other third parties. By following the same direction, we develop the functionalities in Table~\ref{tab:map-gdpr-policies} for each of the GDPR policies as an intermediate representation. This ensures the generated app's functionality complies with GDPR policies.

\begin{table*}[t]
\centering
\small
\caption{Mapping of usage of data with the GDPR policies.}
\begin{tabu}{p{4cm}p{1cm}p{7cm}}
\toprule
    \textbf{Type} & \textbf{Policy}  &  \textbf{Functionality} \\  \toprule
    storage & P1      &  automatically delete data from the app after the retention period \\  \hline
    
  storage, \newline third-party sharing & P2        &  ask for permission whenever collecting personal data from the user end and also ask for permission before sharing data with third-parties   \\  \hline
    
      storage, process, \newline third-party sharing & P3        &  generate GDPR compliant privacy policy using policy generator (Section~\ref{sec:policy-generator}) \\  \hline
    
    storage & P4 &  provide access to individual's data stored on the app \\  \hline
    
    storage &  P5 & allow the users to delete their stored data whenever they want \\  \hline
    
    process  & P6  & encrypt data before sending to the third-parties \\  \hline
    
    storage, process  & P7  & encrypt data before sending to the server \\  
\bottomrule
\end{tabu}
\label{tab:map-gdpr-policies}
\end{table*}

\subsection{Policy Generator} \label{sec:policy-generator} We collect the template of privacy policies from the GDPR website online (\url{https://gdpr.eu/privacy-notice/}). After analyzing the template, we identify information that needs to be consistent with the app functionalities. We provide the modified version of the template policy in the Appendix~\ref{appendix:template-policy}. 
In this version, we highlight the \textit{placeholder} (mark in red) which \sys is responsible for filling. Some information is specific to each company which we leave for the app developer/owner to fill (e.g., company's name, contact number, mailing address). We describe all the description of \textit{placeholders} in Table~\ref{tab:map-purpose-policy-generator}. For example, we will show the list of collected PII in the places marked as `LIST OF PII INFORMATION'. With the help of the previous two components (i.e., information extractor and GDPR policy finder), we determine the values of each of these placeholders listed in Table~\ref{tab:map-purpose-policy-generator} of the template policy. 

\begin{table}[h]
\centering
\small
\caption{Instruction on filling the template policy.}
\begin{tabu}{p{6.7cm}p{9cm}}
\toprule
    \textbf{Placeholder}  & \textbf{Description} \\  \toprule
    PII INFORMATION & collected PII \\ \hline
    THE PROCESS OF PII DATA COLLECTION & list of features which indicate storage type data usage \\ \hline
    PURPOSE & how the app will use the PII data \\ \hline
    ORGANIZATIONS THAT WILL RECEIVE DATA & third party entities which the app will share PII data with \\ 
\bottomrule
\end{tabu}
\label{tab:map-purpose-policy-generator}
\end{table}

\begin{algorithm}[!ht]

    \SetAlgoLined
    \textbf{Input:} seedPurpose \\
    \textbf{Output:} rephrasedPurpose 
    
     rephrasedPurpose = list() \;
     
     \ForEach{$sentence$ $\in$ $seedPurpose$}{
     generatedSentence = paraphrase(sentence)\;
     
     $OIA_{G}$ = OIA(generatedSentence)\;
     
     $OIA_{R}$ = OIA(sentence)\;
     
     $nNodes_R$ = representative(`noun' nodes from $OIA_{R}$)\;
     
     $nNodes_G$ = `noun' nodes from $OIA_{G}$\;
     
     value = checkFluency(generatedSentence)\;
     
     \If{$nNodes_R \subset nNodes_G$ $\&$ value >= $READIBLITY\_THRESHOLD$}{
        add $generatedSentence$ to $rephrasedPurpose$\;
     }
    }
         \caption{Generate purpose for privacy policy}
     \label{algo:generate-diverse-policy}
    \Return rephrasedPurpose

    \end{algorithm}

\newpage

In the privacy policy, an app is expected to state the reason for collecting each PII clearly, which we refer to as `purpose'. We generate the purpose by following Algorithm~\ref{algo:generate-diverse-policy} with the help of an existing paraphrase generator tool~\citep{wieting2017learning} and seed purposes. We collect different policies (from popular apps) for different features, which we use as seed data for the paraphrase generator tool. 
In Table~\ref{tab:policy-generator-mapping}, we list the seed purposes that we use for the paraphrase generator. We create the purposes for all the different features considered in this work. We collect those from the privacy policies of ten most popular mobile apps in the google play store, i.e., Facebook, Instagram, WhatsApp, Google, TikTok, Snapchat, Pinterest, Twitter, Reddit, Skype.
For each sentence inside these policies, we generate ten different paraphrases. Then, we investigate whether the generated sentence preserves the actual meaning or not. Due to the difference in the wordings, the rephrased sentence's meaning may change slightly, resulting in ambiguity. But in the privacy policy, we have to ensure that the readers can clearly understand the purpose of collecting PII. Thus, we measure the quality of the paraphrase sentences using  Algorithm~\ref{algo:generate-diverse-policy}. After generating an OIA graph from the original sentence, we identify the representative noun nodes. Note that a node is defined as a representative noun node when it contains more than two edges (either incoming or outgoing or both) in the OIA graph or PII information. From our investigation, we find these noun nodes are essential in preserving the sentences' actual meaning. If the representative noun nodes are missing in the generated paraphrase graph, we mark that as `not preserving'. To ensure the readability of the generated purpose, we check the readability level of each of the sentences. We calculate three scores for each generated purpose, i.e., Dale-Chall Score~\citep{dale-chall-score}, Gunning's FOG Index~\citep{gunning-fog-score}, and Flesch Reading Ease~\citep{flesch-reading-score}, as described in Section~\ref{sec:readability}. We examine whether the generated purpose is easily understandable by at least a 7th or 8th grade student by matching the corresponding value for each of these scores. Consequently, we create a diverse set of readable GDPR compliant policies.

\begin{table*}[t]
\centering
\small
\caption{Diverse policy generation. Note that, we collected the purposes from 10 most popular mobile applications (including-- Facebook, Instagram, WhatsApp, Google, TikTok, SnapChat, Pinterest, Twitter, Reddit, Skype). Here, `you' refers to the person using the app.}\vspace{0.1in}
\begin{tabu}{p{4cm}p{11.5cm}}
\toprule
     \textbf{Feature}  &  \textbf{Purpose} \\  \toprule
    registration + change password + add new friends   & We collect the content, communications, and other information you provide when you use our products. \\ \hline
    login & Conduct analytics and research on who is using our app and what they are doing, for example, by logging how often you use a particular feature on our app. \\ \hline
    comments & Review your messages to detect activity that poses a risk to the safety of you and our community. \\ \hline
    user profile + news recommendation & To infer additional information about you, such as your age, gender, hobbies, and interests. We use this information to better recommend content to you. \\ \hline
    user status + blog writing + create new post + review + notes + status updates & To fulfill requests for products, services, platform functionality, support, and information for internal operations, including troubleshooting, data analysis, testing, research, statistical, and survey purposes, and to solicit your feedback. \\ \hline
    news feed + home feed & To customize the content that you will see when you use our app. For example, we may provide you with services based on the location or time. \\ \hline
    address book & You can choose to sync your address book on our app so that we can help you find and connect with your friends and help others find and connect with you.  \\ \hline
    people nearby + chat with friends + search for people nearby & Users with whom you communicate can store or share your information (including your phone number) with others on and off our services. \\ \hline
    share + third-party integrations & We may share information about you with third-party services, such as advertising partners, data providers, and analytics providers. \\ \hline
    app purchase + product scan & You may provide us with payment information, including your credit or debit card number, expiration date, CVV, and billing address, in order to purchase our services. If you make a payment using our app, we may receive information about your transaction such as when it was made or when a subscription is set to expire or auto-renew. \\

\bottomrule
\end{tabu}
\label{tab:policy-generator-mapping}
\end{table*}

\newpage

However, if our tool fails to predict no UI elements or features or page information in the description, it will automatically ask the user again to revise the description and provide the refined description by highlighting the specific sentence for revision and providing the reasons (e.g., missing UI element, missing feature, missing page information, vague UI element, vague feature, vague page information). This functionality is easily implemented by using the OIA parsing results. In addition, we plan to create a portal where users can request new features or UI elements for facilitating future development.

Once we extract all the information using \sys, we convert all the functionalities to apk file (which can be installed on Android phones) using MIT app inventor~\citep{mit-app-inventor}. MIT app inventor is a popular tool among app developers. We use this tool to generate an executable apk file for mobile devices running Android OS.

\newpage

\section{Evaluation}
\label{sec:evaluation}

In this section, we present the detailed process of evaluating \sys in generating a GDPR compliant app. The tool suite is implemented as a python prototype (Python 3.7). We seek to understand three research questions: 
\begin{itemize}[topsep=1pt, itemsep=1pt, partopsep=1pt, parsep=1pt, leftmargin=1cm]
    \item[$RQ1$] How does each component of \sys perform? 
    \item[$RQ2$] What is the end-to-end performance of \sys?
    \item[$RQ3$] How much easy to understand the generated policy?
\end{itemize}

We conduct all the experiments (Section~\ref{subsec:evaluation:information-extractor} -~\ref{subsec:evaluation:policy-extractor} for $RQ1$, Section~\ref{subsec:evaluation:end-to-end-evaluation} for $RQ2$, and Section~\ref{subsec:evaluation:policy-extractor} for $RQ3$) on a Desktop PC with 16 GB of RAM and a 3.1 GHz Intel Core i5 processor, running Ubuntu 18.04 LTS. We use all the data from the first phase as the `training data' (66 app), and the second phase as the `test data' (36 app). None of the participants involve in both phases.

\subsection{Evaluation of Information Extractor} \label{subsec:evaluation:information-extractor}
Our information extractor component consists of three sub-components. At first, we evaluate the performance of each sub-component.

\subsubsection{Page Information Extraction}\label{subsec:evaluation:page-extractor}

In the descriptions of 48 apps, we had the information of 396 total pages available. The performance of page information extraction is presented in Table~\ref{tab:eval-page-info-extraction}. It can be seen that by using our tool, we were able to detect 357 current page and 347 transition page information successfully, while missed the current page information 39 times and transition page information 49 times, respectively. In summary, \sys achieves 88.9\% accuracy (90.2\% for current and 87.6\% for transition) in identifying page information.

\begin{table}[ht!]
\centering
\caption{Evaluation of extracting page information.}
\begin{tabular}{cccc}
\toprule
      \textbf{Type} & \textbf{\#Matched}  & \textbf{\#Not Matched} & \textbf{Accuracy} \\  \toprule

      Current & 357 & 39  & 90.2\% \\  \hline
      Transition & 347 & 49 & 87.6\% \\
      
\bottomrule
\end{tabular}

\label{tab:eval-page-info-extraction}
\end{table}

\subsubsection{UI Information Extraction}  \label{subsec:evaluation:ui-extractor}

\begin{table}[b!]
\centering
\caption{Performance evaluation in extracting UI elements.}
\begin{tabular}{ccc}
\toprule
       \textbf{\#Matched}  & \textbf{\#Not Matched} & \textbf{Accuracy} \\  \toprule
       476 & 21 & 95.8\% \\
\bottomrule
\end{tabular}
\label{tab:eval-ui-info-extraction}
\end{table}

In our test set, we have 48 app descriptions that contain interactions with 396 sentences. These descriptions contain 497 UI elements (from 396 sentences) in total. In Table~\ref{tab:eval-ui-info-extraction}, we illustrate the detailed performance of extracting UI elements for our tool. Our tool is able to identify 476 (from 388 sentences) UI elements correctly. It fails to detect 21 UI elements from 8 different sentences.
By investigating these misclassifications, we find that \sys, in rare cases, fails to differentiate the descriptions for the UI elements' position from the entity of UI elements. For example, ``In this page, there is an `input' edittext, and alongside the edittext, there is a `search' imagebutton''. In the second half of the sentence, the participant attempts to describe the position of the imagebutton corresponding to the edittext, where the edittext refers to the one in the first half as opposed to a new edittext. This confuses our tool, such that it creates a new `edittext' element in the UI. But these are some rare cases as \sys is able to achieve 95.8\% accuracy.

\subsubsection{Feature Information Extraction}  \label{subsec:evaluation:feature-extractor}

In this section, we evaluate the performance of our tool from the perspective of extracting feature information. We extract the key information related to each feature from each sentence of the descriptions. So we evaluate the performance of our tool in terms of detecting features based on each description.

\begin{table}[h]
\centering
\caption{Performance evaluation in identifying feature.}
\begin{tabular}{cccc}
\toprule
      \textbf{\#Feature} & \textbf{\#Matched}  & \textbf{\#Not Matched} & \textbf{Accuracy} \\  \toprule

      48 & 44 & 4 & 91.7\% \\  
\bottomrule
\end{tabular}
\label{tab:eval-feature-extraction}
\end{table}

From Table~\ref{tab:eval-feature-extraction}, we can see that \sys is able to identify the majority of features correctly with a 92\% accuracy. For example, ``In my `login' page, it has 2 edittexts and 4 buttons. The edittexts are used to enter `email' and `password'. There is a `button' of `login' in the upper right corner. Once the user clicks the button of `login', it will send the user's `email' and `password' to the server. There are 2 buttons below the edittext to avoid forgetting the account or password. Press the button of `Forgot Password' to jump to the `Forgot Password' page, and press the button of `Use Device Code' to jump to the `Use Device Code' page.'' After analyzing all the sentences in this description, our tool predicts the feature as `login'.

\subsection{Evaluation of GDPR Policy Finder} \label{subsec:evaluation:gdpr-policy-finder}

We evaluate \sys regarding the detection of GDPR policies, as shown in Table~\ref{tab:evaluation-gdpr-policy-finder}. Our GDPR policy finder can successfully identify whether a single app description exhibits the characteristics of storage, process, third-party data sharing, or nothing.

\begin{table}[!ht]
\centering
\caption{Performance evaluation of GDPR Policy Finder.}
\begin{tabular}{cccc}
\toprule
\textbf{Type} & \textbf{\#Matched} & \textbf{\#Not Matched} & \textbf{Accuracy} \\ \toprule

 Storage & 368 & 28 & 92.9\% \\ \hline
 Process & 377 & 19 & 95.2\% \\ \hline 
 Third-party Sharing & 390 & 6 & 98.4\%  \\ 

\bottomrule
\end{tabular}
\label{tab:evaluation-gdpr-policy-finder}
\end{table}

From Table~\ref{tab:evaluation-gdpr-policy-finder}, we can observe that \sys achieves 93\%, 95\%, and 98\% accuracy for detecting storage, process, and third-party sharing, respectively. Overall, our tool correctly detects the 368 storage, 377 processes, and 390 third-party sharing data usages.

\newpage

\subsection{Evaluation of Policy Generator} \label{subsec:evaluation:policy-extractor}
With our paraphrase generator, we generate 300 different purposes for 23 features. After evaluating the purposes, we filter out 136 generated purposes that do not convey the original information of the corresponding purposes. That indicates either they do not have the representative noun nodes or their readability is less than the threshold value. We evaluate the rest 164 generated purposes manually. We find that our tool is successful in generating 149 purposes. Thus, \sys achieves 90.9\% accuracy in generating the purposes of the privacy policies.   

\subsection{End-to-end Validation of \sys} \label{subsec:evaluation:end-to-end-evaluation}

We select 48 app descriptions and perform an end-to-end evaluation of our tool. We manually annotate those apps to the corresponding required policy. For example, if a feature shares a user's `email' with third parties, we mark P2, P3, P6, and P7 as the required policies that the generated feature should comply with. Then, we evaluate each generated feature whether it contains these policies or not.

\begin{table}[!ht]
\centering
\caption{End-to-end evaluation of \sys.}\vspace{0.1in}
\begin{tabular}{cccc}
\toprule
\textbf{Policy} & \textbf{\#Matched} & \textbf{\#Not Matched} & \textbf{Accuracy} \\ \toprule

P1 &   43  &  5  & 89.6\%   \\ \hline
P2 &   43  &  5  & 89.6\% \\ \hline
P3 &   42  &  6  & 87.5\%   \\ \hline
P4 &   43  &  5  & 89.6\%   \\ \hline
P5 &   42  &  6  & 87.5\%   \\ \hline
P6 &   44  &  4  & 91.7\%   \\ \hline
P7 &   44  &  4  & 91.7\%   \\ 

\bottomrule
\end{tabular}
\label{tab:evaluation-end-end-evaluation}
\end{table}

In Table~\ref{tab:evaluation-end-end-evaluation}, we show that the number of listed policies (P1-P7) are correctly identified by using our tool. We can observe from this table that our tool can detect all the seven policies with an average of 90\% accuracy. Since the tool can automatically ask the developer to revise any missing information as described in Section~\ref{sec:policy-generator}, if we assume all the descriptions are revised according to the requirement and no feature or UI element is outside the supported lists of our tool, we expect \sys can achieve much higher accuracy in predicting all the seven policies.
With the help of \sys, we are able to mitigate the hurdle of GDPR compliance. Developers don't need to go through long and complex GDPR policies. It is the responsibility of the \sys to ensure the generated feature complies with GDPR regulations will save the developers time in understanding the proper way of protecting user personal information. The developers of \sys have gone through all the GDPR policies and have participated in several discussions related to GDPR to help implement the corresponding requirement of the GDPR policies for a particular feature. 

\newpage

\section{Case Study}
\label{sec:case-study}

In the section, we present an interesting example where \sys is able to build a GDPR compliant app. The complete description of the app is -- ``In `registration' page, the user only needs to input their email address in `edittext' and hit the `sign up' button. There is a `back' button on the bottom of `sign up' button. After pressing the `back' button it will take them to the `welcome' page.''

\begin{figure*}[!ht]
\centerline{\includegraphics[width=6.5in]{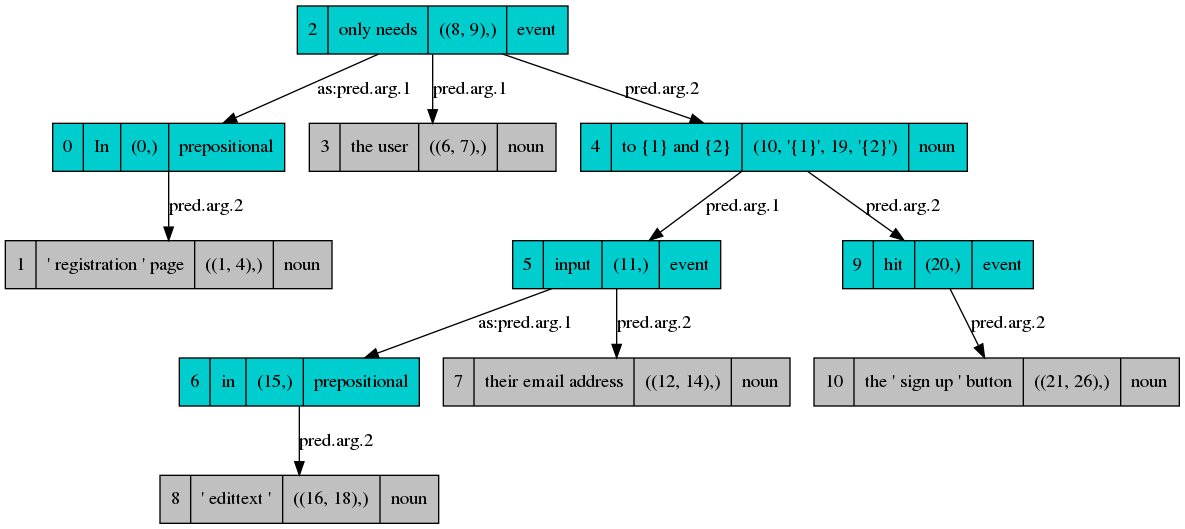}}
\caption{OIA graph of ``In `registration' page, the user only needs to input their email address in `edittext' and hit the `sign up' button''.}
\label{fig-case-study-success}
\end{figure*}

\vspace{0.1in}\noindent\textbf{Information Extractor.}  Our information extractor component is responsible for detecting three key information: page information, UI elements, and feature information. First, it finds the page node information (node id is 1 in Figure~\ref{fig-case-study-success}). Using Algorithm~\ref{algo:page-info-desc}, we detect the current page as `registration'. Second, it can identify all the UI elements (e.g., button, edittext) according to Section~\ref{sec:system_design:ui-information-extractor}. Our tool predicts the feature after analyzing all the sentences from each description. Based on the techniques described in Section~\ref{sec:system_design:feature-information-extractor}, \sys predicts the feature as `registration'. As a result, our information extractor extracts all the necessary information correctly and then provides this to the subsequent component.

\vspace{0.1in}\noindent\textbf{GDPR Policy Finder.} From the generated OIA graph in Figure~\ref{fig-case-study-success}, we can see that it contains `input' as an event node, and the related noun node contains PII (i.e., `email'). From Table~\ref{tab:gdpr-policy-finder}, we can find that if a sentence has `input' as an event and `registration' as a feature, \sys will mark that as `storage' type of data usage. In other words, from the description, our tool is making a prediction that the app will store user email information. According to Section~\ref{sec:background:gdpr-policy-definition}, whenever an app stores PII, all the policies P1-P5, P7 become applicable. Thus, our tool implements a functionality for asking the user's consent whenever there is an edittext that takes `email' as input. Then, before sending data to the server, our tool encrypts all the PIIs. Once the registration is completed successfully, the user will have the option in the app to make a request to access or delete the stored data (`email'). Finally, this component sends the feature information and data usage types to the next component (i.e., Policy Generator), where we use these information to generate privacy policies.

\newpage

\vspace{0.1in}\noindent\textbf{Policy Generator.} \sys generates the purpose for this `registration' feature as ``we will collect the content, communication and other details you provide when you use our products.'' This sentence yields scores of 9.79, 22.53, and 53.21 for the Dale-Chall Score, Gunning's FOG Index, and Flesch Reading Ease, respectively. All these scores are above our pre-defined readability threshold values. By incorporating the previous two components, our tool generates the following purpose: ``we will collect the content, communication, and other details you provide when you use our products.'' The complete privacy policy can be found in the Appendix \ref{appendix:sample-policy}
where all the texts generated by \sys are highlighted in green. We put the default contents in the rest of the placeholders.

\begin{figure*}[h]
\centerline{\includegraphics[width=6in]{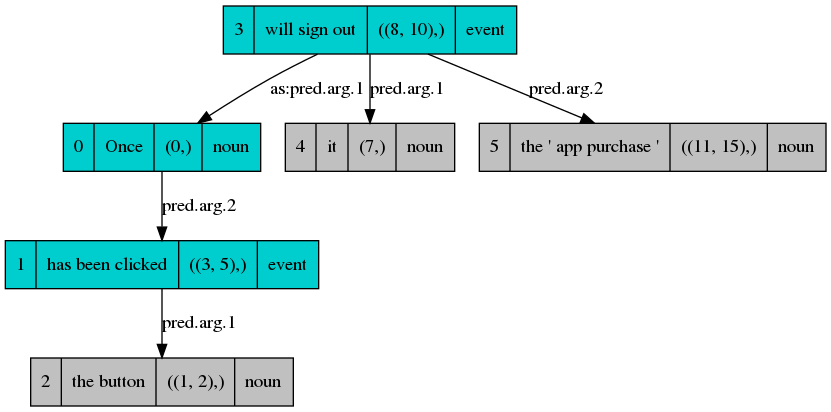}}
\caption{Generated OIA graph for ``Once the button has been clicked, it will sign out the `app purchase'''.}
\label{fig-oia-example-1}
\end{figure*}

\begin{figure*}[h]
\centerline{\includegraphics[width=6.5in]{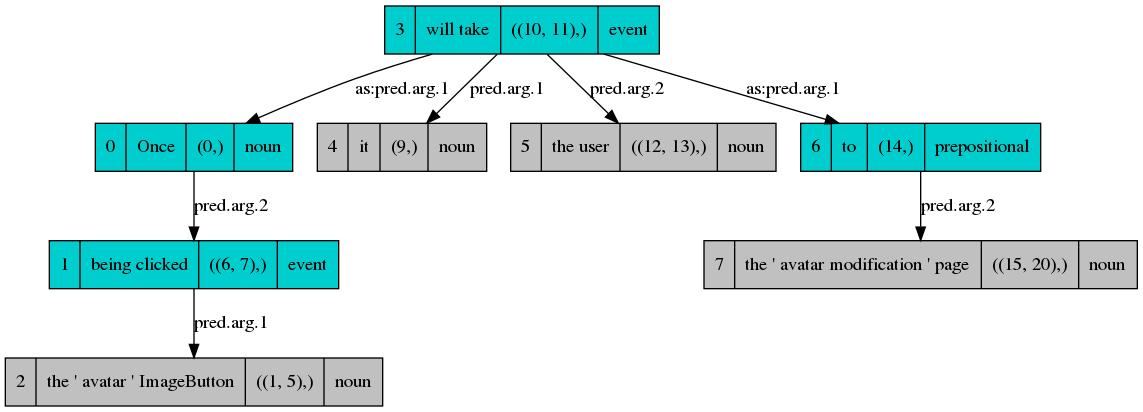}}
\caption{Generated OIA graph for ``Once the `avatar' ImageButton being clicked, it will take the user to the `avatar modification' page''.}
\label{fig-oia-example-3}
\end{figure*}

We present two more examples of OIA graphs in Figures~\ref{fig-oia-example-1} and \ref{fig-oia-example-3} to show that \sys is able to successfully capture the desired information in our privacy-centric task.

\section{Discussion}
\label{sec:discussion}

In this section, we discuss ethical discussion, adaptation to other privacy policies,  additional request submission, future work, and limitations.

\vspace{0.1in}\noindent\textbf{Ethical Considerations.} \label{sec:discussion:ethical-discussion} We are aware of the privacy implications of our work. First of all, our tool is not 100\% accurate. But we put our best effort into improving the tool and detect the required GDPR policy correctly. We preliminarily evaluated the performance of our tool by conducting surveys with computer science students who are familiar with GDPR. According to them, \sys can successfully create a GDPR compliant app for over 90\% of the time, which is also consistent with our end-to-end evaluation. In the future, we plan to contact law experts and will ask them for their feedback. Second, we perform a study to collect app description data from developers. 
Each of the participants is required to have at least two years of development or extensive user experience with mobile applications. More details are discussed in Section~\ref{sec:data-collection}. In our studies, the participants are only asked to provide some functional descriptions of different apps. We have not collected or stored any of their PII (e.g., name, age, gender, IP address, email). The participants are compensated with \$2.33 (paid in a different currency and met the local minimum wage requirement) for each app description.

\vspace{0.1in}\noindent\textbf{Adaptation to Other Privacy Policies.} 
There are several different privacy policies implemented in the different regions of the US. Some of the other popular ones include CCPA (California Consumer Privacy Act), CDPA (Virginia Consumer Data Protection Act), Colorado Privacy Act (CPA). All of them have the exact definition of PII. Besides, they all have a similar requirement like GDPR with only very minor modifications (e.g., retention period varies across different policies). So the generated from our tool can also be easily adapted for compliance with other privacy policies.

\vspace{0.1in}\noindent\textbf{Additional Request.} 
As discussed in Section~\ref{sec:policy-generator}, to accommodate continuous development of \sys with expanded lists of features and UI elements, we plan to create an online development portal where users/developers can submit their requests for adding new features and UI elements. After review, we will develop and integrate these into our list in future releases based on the priority.

\vspace{0.1in}\noindent\textbf{Future Work \& Limitations.} We also want to address a few limitations of our current tool here. 
First, a rule-based approach generally suffers from a lack of coverage, which might lead to false negatives due to insufficient rules. However, from Table~\ref{tab:evaluation-end-end-evaluation}, we can observe that our tool achieves a decent performance with accuracy laying between 86\%-92\% for seven different policies. 
Second, we are only focusing on android app development. But our tool is also extendable (with a minor modification) to generate code for other mobile platforms (such as iOS, native development, etc.). 
Third, we are focusing on an important subset of GDPR policies to build the mobile app. We leave it as a future work for incorporating other GDPR policies as well.
Fourth, we are only focusing on building GDPR compliant mobile apps, while all the server-side computation is beyond the scope of this paper. For example, whether the developer actually deletes the personal data from the server is out of the scope of this paper. An interesting follow-up work entails making APIs and server-side implementation, and consequently the entire mobile development ecosystem, to be also compliant with GDPR. 

\newpage

\section{Conclusion}
\label{sec:conclusion}

We proposed \sys, the first automatic policy generation tool for an app to create GDPR compliant policies from natural language descriptions. It is designed to assist app developers in ensuring compliance with GDPR for the generated app, which will significantly alleviate the burdens for the developers. To build \sys, we adapted an information extraction tool, OIA, and developed several optimization techniques for our privacy-centric task \citep{sun2020predicate,wang2022oie}. By using the extracted information from the natural language descriptions, \sys finds the associated GDPR policy and generates compliant privacy policies. The effectiveness and advantages of \sys are comprehensively evaluated, which shows that \sys can achieve superior performance in detecting various GDPR policies. Our results confirm \sys can successfully generate GDPR compliant policies and functionalities for the app.

\bibliographystyle{plainnat}
\bibliography{refs_scholar}



\section{Appendix}

In this section, we have illustrated template policy and one sample policy generated by our tool.

\subsection{Template Policy}
\label{appendix:template-policy}

[COMPANY] is part of the [COMPANY] Group which includes [COMPANY]
International and [COMPANY] Direct. This privacy policy will explain how our
organization uses the personal data we collect from you when you use our website.
Topics:
\begin{itemize}
\item  What data do we collect?
\item  How do we collect your data?
\item  How will we use your data?
\item  How do we store your data?
\item  Marketing
\item  What are your data protection rights?
\item  Privacy policies of other websites
\item  Changes to our privacy policy
\item  How to contact us
\item  How to contact the appropriate authorities

\end{itemize}

\noindent What data do we collect?
[COMPANY] collects the following data:

\begin{itemize}
    \item LIST OF PII INFORMATION
\end{itemize}

\noindent How do we collect your data?
You directly provide [COMPANY] with most of the data we collect. We collect data
and process data when you:

\begin{itemize}
    \item LIST OF THE PROCESS OF PII DATA COLLECTION
\end{itemize}

\noindent How will we use your data?
[COMPANY] collects your data so that we can:

\begin{itemize}
    \item PURPOSE
\end{itemize}

\newpage

\noindent If you agree, Our Company will share your data with our partner companies so that
they may offer you their products and services.

\begin{itemize}
\item LIST ORGANIZATIONS THAT WILL RECEIVE DATA
\end{itemize}
When Our Company processes your order, it may send your data to, and also use
the resulting information from, credit reference agencies to prevent fraudulent
purchases.
How do we store your data?
Our Company securely stores your data and we will process your data by following
standard security protocol.
[COMPANY] will keep your [LIST OF PII INFORMATION] for [TIME PERIOD]. Once
this time period has expired, we will delete your data.
Marketing
[COMPANY] would like to send you information about products and services of
ours that we think you might like, as well as those of our partner companies.

\begin{itemize}
\item LIST ORGANIZATIONS THAT WILL RECEIVE DATA
\end{itemize}

If you have agreed to receive marketing, you may always opt out at a later date.
You have the right at any time to stop [COMPANY] from contacting you for
marketing purposes or giving your data to other members of the Our Company
Group.
If you no longer wish to be contacted for marketing purposes, please click here.
What are your data protection rights?
[COMPANY] would like to make sure you are fully aware of all of your data
protection rights. Every user is entitled to the following:
The right to access – You have the right to request [COMPANY] for copies of your
personal data. We may charge you a small fee for this service.

The right to rectification – You have the right to request that [COMPANY] correct
any information you believe is inaccurate. You also have the right to request
[COMPANY] to complete the information you believe is incomplete.
The right to erasure – You have the right to request that [COMPANY] erase your
personal data, under certain conditions.
The right to restrict processing – You have the right to request that [COMPANY]
restrict the processing of your personal data, under certain conditions.
The right to object to processing – You have the right to object to [COMPANY]
processing of your personal data, under certain conditions.
The right to data portability – You have the right to request that [COMPANY]
transfer the data that we have collected to another organization, or directly to you,
under certain conditions.
If you make a request, we have one month to respond to you. If you would like to
exercise any of these rights, please contact us at our email:
Call us at: [PHONE NUMBER]
Or write to us: [MAILING ADDRESS]
Privacy policies of other websites
The [COMPANY] website contains links to other websites. Our privacy policy
applies only to our website, so if you click on a link to another website, you should
read their privacy policy.
Changes to our privacy policy
Our Company keeps its privacy policy under regular review and places any updates
on this web page. This privacy policy was last updated on [DATE].
How to contact us
If you have any questions about [COMPANY] privacy policy, the data we hold on
you, or you would like to exercise one of your data protection rights, please do not
hesitate to contact us.

Email us at: [EMAIL]
Call us: [PHONE NUMBER]
Or write to us at: [MAILING ADDRESS]
How to contact the appropriate authority
Should you wish to report a complaint or if you feel that Our Company has not
addressed your concern in a satisfactory manner, you may contact the Information
Commissioner's Office.
Email: [ADMIN\_EMAIL]
Address: [ADMIN\_ADDRESS]

\newpage

\subsection{Sample Policy}
\label{appendix:sample-policy}

[MY SAMPLE APP] is part of the [MY SAMPLE APP] Group which includes [MY SAMPLE APP]
International and [MY SAMPLE APP] Direct. This privacy policy will explain how our
organization uses the personal data we collect from you when you use our website.
Topics:
\begin{itemize}
 \item What data do we collect?
 \item How do we collect your data?
 \item How will we use your data?
 \item How do we store your data?
 \item Marketing
 \item What are your data protection rights?
 \item Privacy policies of other websites
 \item Changes to our privacy policy
 \item How to contact us
 \item How to contact the appropriate authorities
\end{itemize}

\noindent What data do we collect?
[MY SAMPLE APP] collects the following data:
\begin{itemize}
    \item Email
\end{itemize}

How do we collect your data?
You directly provide [MY SAMPLE APP] with most of the data we collect. We collect data
and process data when you:
\begin{itemize}
    \item Registration
\end{itemize}

How will we use your data?
[MY SAMPLE APP] collects your data so that we can:
\begin{itemize}
    \item we will collect the content, communication and other details you
provide when you use our products.
\end{itemize}

If you agree, Our Company will share your data with our partner companies so that
they may offer you their products and services.
 \begin{itemize}
    \item  None
\end{itemize}

When Our Company processes your order, it may send your data to, and also use
the resulting information from, credit reference agencies to prevent fraudulent
purchases.
How do we store your data?
Our Company securely stores your data and we will process your data by following
standard security protocol.
\begin{itemize}
    \item  MY SAMPLE APP will keep your [Email] for [365 days]. Once this time period has
expired, we will delete your data.
\end{itemize}
Marketing
[MY SAMPLE APP] would like to send you information about products and services of
ours that we think you might like, as well as those of our partner companies.
\begin{itemize}
    \item  List organizations that will receive data [None]
\end{itemize}

If you have agreed to receive marketing, you may always opt out at a later date.
You have the right at any time to stop [MY SAMPLE APP] from contacting you for
marketing purposes or giving your data to other members of the Our Company
Group.

If you no longer wish to be contacted for marketing purposes, please click here.
What are your data protection rights?
[MY SAMPLE APP] would like to make sure you are fully aware of all of your data
protection rights. Every user is entitled to the following:\\

\para{The right to access} You have the right to request [MY SAMPLE APP] for copies of your
personal data. We may charge you a small fee for this service.

\para{The right to rectification} You have the right to request that [MY SAMPLE APP] correct
any information you believe is inaccurate. You also have the right to request
[MY SAMPLE APP] to complete the information you believe is incomplete.

\para{The right to erasure} You have the right to request that [MY SAMPLE APP] erase your
personal data, under certain conditions.

\para{The right to restrict processing} You have the right to request that [MY SAMPLE APP]
restrict the processing of your personal data, under certain conditions.

\para{The right to object to processing} You have the right to object to [MY SAMPLE APP]
processing of your personal data, under certain conditions.

\para{The right to data portability} You have the right to request that [MY SAMPLE APP]
transfer the data that we have collected to another organization, or directly to you,
under certain conditions.

If you make a request, we have one month to respond to you. If you would like to
exercise any of these rights, please contact us at our email:
Call us at: [1112223333]
Or write to us: [MY SAMPLE APP@email.com]
Privacy policies of other websites
The [MY SAMPLE APP] website contains links to other websites. Our privacy policy applies
only to our website, so if you click on a link to another website, you should read
their privacy policy.
Changes to our privacy policy
Our Company keeps its privacy policy under regular review and places any updates
on this web page. This privacy policy was last updated on 20 July 2021.
How to contact us
If you have any questions about [MY SAMPLE APP] privacy policy, the data we hold on
you, or you would like to exercise one of your data protection rights, please do not
hesitate to contact us.

Email us at: [MY SAMPLE APP@email.com]

Call us: [1112223333]

Or write to us at: [MY SAMPLE APP@email.com]

How to contact the appropriate authority?

Should you wish to report a complaint? Or if you feel that Our Company has not
addressed your concern in a satisfactory manner, you may contact the Information
Commissioner's Office.

Email: [MY SAMPLE APP@email.com]

Address: [MY SAMPLE APP@email.com]

\end{document}